\documentclass[english]{pfia}

\usepackage{url}
\usepackage{breakurl}
\usepackage[breaklinks,hidelinks]{hyperref}
\usepackage[numbers]{natbib}

\usepackage{graphicx}

\title{\textbf{A Review of Validation and Verification of Neural Network-based Policies for Sequential Decision Making}}

\newcommand{\fup}[1]{\textsuperscript{#1}}
\newcommand{\citeall}[1]{\citeauthor{#1} (\citeyear{#1}) \cite{#1}}

\author{Q. Mazouni\fup{1}, H. Spieker\fup{1}, A. Gotlieb\fup{1}, M. Acher\fup{2}\\[6pt]
\fup{1} Simula Research Laboratory, Oslo, Norway\\
\fup{2} Univ Rennes, Inria, INSA Rennes, CNRS, IRISA, France
}

\date{quentin@simula.no}

\begin{document}

\maketitle


\begin{resume}
Les réseaux de neurones sont aujourd'hui couramment utilisés pour représenter et apprendre la stratégie des agents pour la prise de décision séquentielle. 
Ce domaine d'application implique de nouveaux défis d'évaluation de la qualité des logiciels que les pratiques de validation et de vérification traditionnelles ne sont pas en mesure de résoudre. 
En conséquence, des approches novatrices ont émergé pour adapter ces techniques aux stratégies basées sur les réseaux de neurones pour la prise de décision séquentielle.
Ce document vise à résumer ces nouvelles contributions et à proposer de futures directions de recherche.
Nous avons effectué une revue de la récente littérature (de 2018 à 2023), dont les sujets couvrent des aspects du test ou de la vérification des stratégies basées sur les réseaux de neurones.
La sélection des travaux a de plus été enrichie par un processus boule de neige à partir de ceux précédemment sélectionnés, afin d'étendre la portée de cette étude et de fournir au lecteur des informations sur les défis de vérification similaires et leurs récentes solutions. 
Finalement, nous avons selectionné 18 articles.
Nos résultats témoignent d'un intérêt croissant pour cette problématique. 
Ils mettent en évidence la diversité des problèmes exacts considérés et des techniques utilisées pour y faire face.
\end{resume}

\begin{motscles}
Test, Réseau de neurones, Prise de décision séquentielle.
\end{motscles}

\begin{abstract}
In sequential decision making, neural networks (NNs) are nowadays commonly used to represent and learn the agent's policy.
This area of application has implied new software quality assessment challenges that traditional validation and verification practises are not able to handle. 
Subsequently, novel approaches have emerged to adapt those techniques to NN-based policies for sequential decision making.
This survey paper aims at summarising these novel contributions and proposing future research directions.
We conducted a literature review of recent research papers (from 2018 to beginning of 2023), whose topics cover aspects of the test or verification of NN-based policies. 
The selection has been enriched by a snowballing process from the previously selected papers, in order to relax the scope of the study and provide the reader with insight into similar verification challenges and their recent solutions. 
18 papers have been finally selected.
Our results show evidence of increasing interest for this subject. 
They highlight the diversity of both the exact problems considered and the techniques used to tackle them.
\end{abstract}



\section{Introduction}

Last years have seen tremendous advances in solving sequential decision making problems with neural networks (NNs). For example, in game playing \cite{article, doi:10.1126/science.aar6404} or Artificial Intelligence (AI) Planning \cite{DBLP:journals/corr/abs-1908-01362, DBLP:journals/corr/abs-2007-06702, Issakkimuthu_Fern_Tadepalli_2018}. 
Testing software systems that employ these NNs as policies is difficult and encounters several novel challenges.
On one hand, most of the work on NN verification focus on single calls (like in image classification), without thus accounting for the sequential decisions made by the policy tested. 
On the other hand, traditional validation and verification (V\&V) techniques for sequential decision making must deal with the major paradigm shift involved by the use of NNs, where the logic of the program has been learned rather than ``coded''. 
In fact, they can either leverage the subsequent, specific information of NN white-box testing (like NNs' architecture and weights) or assume no information at all about the NN-based policy under test (i.e, black-box testing).
Fortunately, a new research area dedicated to the V\&V of NN-based policies has emerged. 
However, its contributions are very diverse: they don't share the same testing goals and assumptions, and their respective limitations are unclear.


This paper aims at covering this recent literature (from 2018 to beginning of 2023), by detailing all the sub-problems addressed, the methodologies used and their current limitations. 
To cope with the diverse nature of the contributions and their respective degree of maturity, we categorise them (methodology-wise) and include in their review insight into their applicability. 
Furthermore, we highlight the subsequent remaining challenges and suggest possible ideas to address them.
Our purpose is to provide the reader with a succinct, yet comprehensive view of this evolving area of research, in order to stimulate the latter and guide interested researchers towards the uncovered problems. 
We thus address the following questions:
\begin{itemize}
\item What approaches have been recently proposed to address the V\&V of NN-based policies?
\item What are remaining gaps with respect to the V\&V of NN-based policies?
\end{itemize}

Other works have reviewed related topics with different methodologies than ours.
\citeall{10.1613/jair.1.12716} cover autonomous cyber-physical systems (CPSs) rather than NN-based policies. Closer to this work, \citeall{ZHANG2020106296} review NN-based CPSs. However, they propose a systematic literature review (SLR), which drastically differs from our scope and methodology. Besides, it does not include the most recent contributions (the papers were gathered from 2011 to 2018). Eventually, \citeall{10.1007/s10515-022-00337-x} propose another SLR, that aims at answering the question of the certification of learned-based safety-critical systems. Therefore, their study adopts a broader approach than ours and, as \cite{ZHANG2020106296}, lacks the very last contributions this review covers.

We recognise that our survey is not exhaustive and that it does not cover all the literature directory (as the previously aforementioned SLRs do).
Instead, we aim at analysing a more restricted and precise research topic, which is the V\&V of NN-based policies for sequential decision making.

The rest of the paper is organised as follows. Section \ref{background} introduces the relevant notions and concepts that are discussed throughout this paper.
We then describe our paper search and selection methodology in Section \ref{methodology}, as well as statistical results over the 18 papers analysed. 
Section \ref{review} details the review of the papers.
In Section \ref{limitations}, we synthesise the observed limitations. 
Section \ref{conclusion} concludes this paper by elaborating on future research directions. 

\section{Background}\label{background}
In this section, we introduce the key concepts which are at the core of this study: sequential decision making and neural networks as policies.

\subsection{Sequential Decision Making}

Informally, sequential decision making refers to tasks that can be solved by any decision theory in a step by step manner and which accounts for the dynamics of the environment \cite{frankish_ramsey_2014}.
In our study, we consider goal-oriented sequential decision making problems, where an agent starting from an initial state of the world can interact with the environment (e.g, simulations) through step-wise observation-decision-action processes until a satisfying state is reached. 
A typical example is the case of path planning in Robotics, where the agent is expected to safely reach a given position from an initial situation. 
The papers studied in this survey formalise sequential decision making problems as Markov Decision Processes (MDPs).
It is defined as 4-tuple $\langle S,A,R,P\rangle$ where:
\begin{itemize}
    \setlength{\parskip}{5pt}
    \setlength{\itemsep}{0pt plus 1pt}
    \item $S$ is a set of states. Referred as observation space, it specifies what the agent can know about its environment.
    \item $A$ is the set of actions. Referred as action space, it specifies how the agent can act on its environment. 
    \item $R$ is the reward function. It reflects the agent’s performance by associating any pair of state-action with a numerical value. In goal-oriented problems, such functions are often sparse, meaning that the agent receives positive rewards only for goal states (0 otherwise).
    \item $P$ is the transition function, which is a probability distribution over the observation and the action space. It depicts which state the environment will transit to after an action is executed.
\end{itemize}
Solutions to MDPs are called policies (noted $\pi$), which are functions mapping from the observation space $S$ to the action space $A$.

\subsection{Neural Networks as Policies}

The papers studied in this review consider policies as NNs. 
In such a context, the inputs of the networks are usually the observation space of the MDP of a decision making problem (or a slightly adapted version), while their outputs describe a probability distribution over the possible actions. 
Consequently, an agent following a policy $\pi$ means that at every time step $t$, it chooses the next action $a_{t+1}$ whose probability is the highest, i.e, $a_{t+1} = \arg \max \pi(s_t)$). 
Furthermore, we introduce stateless and stateful agents, since some contributions specifically target one or the other. 
Stateless agents follow policies modeled by feed forward neural networks (FFNNs), which consist in multiple hidden layers and admit no cycles \cite{Abiodun2018StateoftheartIA}.
On the other hand, stateful agents rely on recurrent neural networks (RNNs), whose ability to employ sequential data let them recall information \cite{Mandic2001RecurrentNN} (thus providing the agent with a ``memory"). Note that such stateful problems have an extended definition compared to the one defined above. Precisely, the observation space is a set of states (i.e, the current and past observations).

\section{Methodology}\label{methodology}

In this section, we first elaborate on the selection strategy of papers for our literature review. Then, we provide a succinct statistical analysis of the papers selected.

\subsection{Paper Search Strategy}

Since the topic of this survey is quite narrow, we did not conduct a keyword-based, automated search in digital libraries (as systematic literature reviews do). 
We adopted an iterative process instead, whose steps would increasingly enlarge the scope of the search. 
That way we were able to precisely control, for each paper, the relation between its contribution and the topic of the study, as well as monitor the total number of papers. 
Each iteration consisted in a traditional, two-step process where a first batch of papers is reviewed and then we snowballed from their references. We sourced the papers in the first step from the 2022 edition of highly ranked conferences interested in either decision making and AI (ICAPS, IJCAI, AAAI) or software testing and engineering (ICSE, ISSTA, ESEC/FSE, IEEE TSE) and looked for terms related to V\&V practices (e.g, testing, verification).
In addition, we examined papers reviewed in other related surveys \cite{10.1613/jair.1.12716, ZHANG2020106296, 10.1007/s10515-022-00337-x} whose topics match the scope of our study.

\subsection{Selection Criteria}

For each paper gathered during our iterative search, we looked at its title, keywords (if any) and read the abstract. We then completed the reading if (i) the abstract described interests in the validation and verification of programs for solving any decision making problem and (ii) whose assumptions correspond to the ones of NN-based policies (e.g, any learned models, black-box agents). 

We adopted loose quality selection criteria. Indeed, the research topic studied is very active and, since we aim at providing the reader with as many diverse approaches as possible, we argue that only considering peer-reviewed papers would not have let us fulfill this goal.
Therefore, we considered preprints too.

\subsection{Statistical Results}

\begin{figure}[t]
\centerline{\includegraphics[width=\columnwidth]{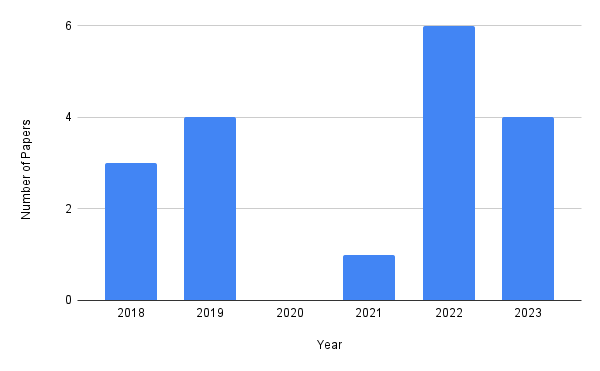}}
\caption{Number of analysed papers per year.}
\label{yeardistribution}
\end{figure}

\begin{figure}[th]
\centerline{\includegraphics[width=\columnwidth]{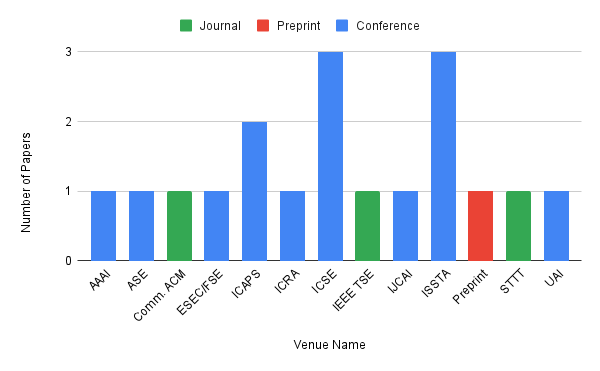}}
\caption{Venue distribution of the selected papers. The venue types are indicated with different colors.}
\label{venuedistribution}
\end{figure}

Figure~\ref{yeardistribution} shows the distribution of the papers' year of publication. 
We can see that most of the papers have been published in the last three years. 
Actually, more than half of them have been published either in 2022 or early 2023. 
This observation highlights how V\&V of NN-based policies has recently become popular. Besides, Figure~\ref{venuedistribution} depicts the venues represented by our paper selection along with their type. 
14 papers were published in conferences, 3 in journals and 1 is a preprint.

\section{Review}\label{review}

\begin{figure*}[t]
\centerline{\includegraphics[width=\textwidth]{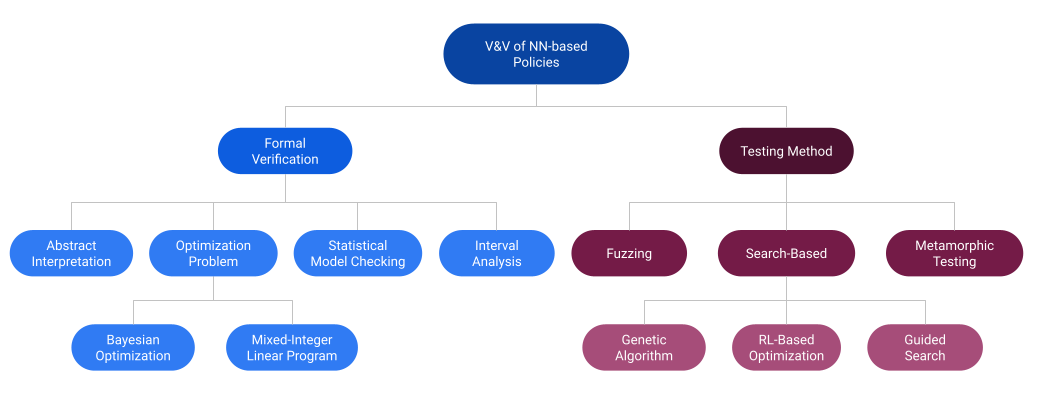}}
\caption{Representation of the taxonomy of the papers reviewed. Each leaf denotes a software testing approach leveraged by at least one of the works presented.}
\label{taxonomy}
\end{figure*}

We derived from our literature review a set of categories that we use to classify the 18 papers selected.
Figure~\ref{taxonomy} shows the result as a comprehensive taxonomy tree, where each leaf denotes a software V\&V technique that best describes the general approach used by at least one of the papers selected.
In the remainder of the section, we present each paper by describing its contribution, summarising its assumptions and highlighting its limitations.

\subsection{Formal Verification Methods}

In our context, verification methods (``Formal Verification'' node on the left side of Figure~\ref{taxonomy}) aim at proving that the agent following the policy under test is safe with respect to safety properties.
These methods return SAT if the specifications are always satisfied (i.e the policy is verified) or UNSAT -- with the associated counterexample -- if they are not.
A counterexample can for example be an entire execution trace of the agent (interacting with the environment) or a state of the world the agent led the simulation to. 

\subsubsection{Statistical Model Checking}

Statistical Model Checking (SMC) \cite{10.1007/978-3-540-24622-0_8} is an alternative to Model Checking \cite{CLARKE20011635} that aims at alleviating the well-known state explosion problem by combining simulation and statistical methods to provide statistical evidence for the satisfaction or violation of the specification.
In essence, the model under test is simulated to generate samples, which are then evaluated with respect to a given property. As such, SMC provides an estimation of the value of the property, along with statistical results on the potential error.
\citeall{Gros2022} propose Deep Statistical Model Checking (DSMC), where the NN-based policy under test is used as an action oracle to select the actions to perform during the execution of the MDP model of the problem. 
In other words, the verification is done through statistical model checking of the MDP whose transitions follow the policy tested. 
The authors evaluate their approach on a simplified version of the Racetrack benchmark\footnote{\url{https://racetrack.perspicuous-computing.science/}}, an autonomous driving challenge, where the objective is to reach the goal in a minimal number of steps without bumping into a boundary wall. 
Interestingly, DSMC has already been used to improve reinforcement learned policies under test \cite{10.1007/978-3-030-85172-9_11}, and has been integrated inside a toolbox, called MoGym \cite{10.1007/978-3-031-13188-2_21}, which enables the training along with the verification of machine-learned agents in an unified framework.
In conclusion, this methodology verifies black-box NN-based policies, but requires a formal (and executable) model of the decision making problem (current implementation uses JANI \cite{10.1007/978-3-662-54580-5_9} models).

\subsubsection{Abstract Interpretation}

One of the abstract interpretation techniques for formal verification consists in checking the specification against an abstract model which over-approximates the concrete model of the system under test.
Thanks to such an over-approximation, the satisfaction of the property by the abstract model also proves the initial model's correctness.
\citeall{Vinzent_Steinmetz_Hoffmann_2022} use predicate abstraction \cite{10.1007/3-540-63166-6_10} to compute a policy abstract state space and checks that none of its states violates a given safety property.
They evaluate their approach with problems taken from the AI Planning literature, adapted to include unsafety conditions and non-deterministic actions.
The results show that their framework outperforms standard predicate abstraction (thus ignoring the policy) and is more applicable than explicit enumeration and bounded model checking baselines.

This methodology has a limited applicability, as it requires a white-box model of the NN tested, as well as a formal model of the decision making problem. More importantly, the technique inputs the abstract predicates (i.e, they are not computed automatically), which significantly increases the amount of testing efforts. 
Besides, most of the optimizations studied only apply to NNs whose activation functions are rectified linear unit (ReLU).

\subsubsection{Interval Analysis}

Formal verification of NNs with reachability methods \cite{8431048} leverages interval analysis \cite{10.5555/904631} to compute and analyse the possible output sets of each layer of the
network.
However, the usual definition of safety properties does not allow their verification in the case of NN-based policies, where the networks' outputs typically encode a probability distribution over the actions. 
\citeall{pmlr-v161-corsi21a} extend this approach to consider the multiple outputs of NN-based policies by introducing behavioral properties, and propose to deal with huge input state spaces of decision making problems (i.e, all the possible situations of the world) by computing an iterative bisection of the input intervals.
In essence, their methodology consists in splitting the input space into areas for which the outputs' boundaries of the NN-based policy never overlap (i.e, the policy can be unambiguously evaluated).
By doing so, the proposed framework is able to quantify the number of violations, from which a violation rate is derived (as the percentage of the input area that causes a violation). 
This metric brings better insight into how the policy performs with respect to the given properties (than the usual SAT/UNSAT output of verification procedures). 
Besides, their implementation, called ProVe, takes advantage of the computation independence of the intervals to parallelize the process.
Consequently, the experimental evaluation conducted shows significant performance improvements over state-of-the-art NN verification tools.
In conclusion, this technique does not need to know the dynamics of the decision making problem, but requires the NN-based policy tested to be a white-box and the safety properties have to be translated in behavioral ones.

\subsubsection{Optimization Problem}
The formal verification task can be seen as an optimization problem, where the verification procedure aims at finding a counterexample as fast as possible.

\paragraph{Bayesian Optimization.}
\citeall{DBLP:journals/corr/abs-1802-08678} encode safety properties as constraints and use Bayesian Optimization (BO) to solve the problem. To do so, the authors first compute a parse tree of the properties and estimate confidences of the lower bound values of the predicates with Gaussian Process. Then, they search for a counterexample in a active testing loop by iteratively minimising the predicates' variables through BO. 
The key idea is that the aforementioned variables are actually parameters of the environment, so each new environment selected minimizes the worst case prediction of violating the properties.
As a result, the exhaustive search is effectively guided towards adversarial environments.
This methodology considers both the simulator and the NN-based policy as black-boxes, but requires a bound value on the agent's trajectories (i.e, bounded verification).

\paragraph{Mixed-Integer Linear Program.} \citeall{Akintunde_Kevorchian_Lomuscio_Pirovano_2019} tackle the case of ReLU-RNN-based policies (i.e, stateful policies) by unrolling the neural network to enable the use of existing white-box verification techniques for FFNNs.
The subsequent problem is then solved with Mixed-Integer Linear Programming (MILP).
They implement their approach in a tool called RNSVerify and compare two unrolling methods, namely: Input on Start (IOS) and Input on Demand (IOD).
Experimental results show that IOD performs systematically better than IOS, since the number of variables and constraints are lower.  
In conclusion, this first research effort for the verification of stateful agents requires access to both the NN and the model of the problem.
As for noticeable limitations, one can remark that such a model has to be linearly-definable (or linearly approximated), the exhaustive search is bounded (like \cite{DBLP:journals/corr/abs-1802-08678}) and the experimental results revealed scalability issues.

\subsection{Testing Methods}

Testing methods (referred as ``Testing Method'' on Figure~\ref{taxonomy}) aim at generating test cases to probe the quality of the policy under test with respect to evaluation criteria. The exact form and meaning of the test cases generated vary from paper to paper, as well as the definitions of the criteria. 
In any case, such criteria rely on the availability of an oracle (i.e, the expected correct output for a given input), which can be explicit translations of safety properties, computed automatically during testing (e.g, metamorphic testing \cite{Chen1998}) or based on more advanced testing techniques like differential testing \cite{McKeeman1998DifferentialTF}.

\subsubsection{Metamorphic Testing}

\citeall{10.1145/3238147.3238187} test image input-based NNs for autonomous driving systems to detect behavior inconsistencies with metamorphic testing (MT) \cite{Chen1998}.
MT is a testing technique that replaces test oracle checking with metamorphic relations (MRs) which assess the results of multiple program executions by specifying how given changes to an input should affect the output. 
As such, MR definitions are usually based on properties of the algorithms implemented. For example, a program that adds $a$ and $b$ should return the same result for the inputs $(a, b)$ and $(b, a)$ -- whatever the actual result is -- since the addition function is commutative.

In this work, the MRs induce weather-based scene changes which are assumed to keep the original semantic for the neural network (i.e, its outputs should not change). 
These transformations are generated by a model that learns to compute different versions (e.g, rainy, snowing) of a single input. 
To do so, the model combines a generative adversarial network and a variational autoencoder. 
In their experimental evaluation, their implementation -- called DeepRoad -- show better synthetic image transformations compared to DeepTest \cite{10.1145/3180155.3180220} (a competitive approach, described in \ref{coverageguidedsearch}), but this comes with the cost of training the aforementioned model first.
In particular, such a training requires pair-set data collections, where images of same driving situations under different weather conditions have to be regrouped together.

\subsubsection{Search-Based Testing}

In the following, we analyse contributions whose overall frameworks are inspired by Search-Based Software Testing \cite{5954405}.
In general, search-based approaches consist in searching for input test cases and gathering the ones whose outputs reveal wrong behaviors of the policy under test. 
Similarly to other testing methods, erroneous behaviors are detected thanks to test oracles.
Since the search space is in most cases very large, the challenge is to develop optimizing techniques to efficiently find fault-revealing test cases within the available test budget.

\paragraph{Genetic Algorithm.} \citeall{zolfagharian2023searchbased} and \citeall{9794023} optimize the search with genetic algorithms \cite{10.2307/24939139}.
Such algorithms consider test cases as individuals of a population. The general idea is to iteratively let the population evolve (through crossover and mutation transformations) and only retain the most promising individuals (with a selection function) for the next iteration. Therefore, a typical genetic algorithm repeats the following until the test budget is consumed: (i) generating a new set of individuals with the crossover and mutation operators from the current population; (ii) calculating their fitness scores by executing the policy for every individual; (iii) keeping the test cases which revealed wrong behaviors (given a test oracle); (iv) selecting the individuals of the population for the next iteration with the selection function.

\citeauthor{zolfagharian2023searchbased}~\cite{zolfagharian2023searchbased} test the policy of Reinforcement Learning (RL) based agents in a data-box testing setting (i.e, the training data is accessible) by finding faulty episodes.
As such, the individuals of the population are episodes, and their genes are the state-action pairs of the execution traces of the latter.
An abstract representation of the observation space (of the MDP model of the problem) is first computed with the aforementioned training data, which then lets their framework reason over abstract states to combine individuals of the population.
Precisely, the combination of two episodes consists of the genes of the first individual but whose last genes, starting from a crossover point (randomly selected), are replaced by the ones of the second individual. The authors best preserve the consistency of the new episode (i.e, it can be executed by the RL agent) by checking the two concrete states designated by the crossover point belong to the same abstract class.
They define three fitness functions, which favor low-reward episodes, episodes that maximise policy's uncertainty level and the ones that minimise the probability of functional faults, respectively. 
The probability of those faults is predicted with Machine Learning (Random Forest) whose model is trained in an initial step of the methodology with the training data of the agent under test. 

In addition to the restricted scope of this work (it only applies to RL agents), we point out several weaknesses. 
First, the authors had a hard time ensuring the consistency or realism of the faulty episodes, since they are mutated. Furthermore, the validity of the test cases is only taken into account when they are finally compared with the execution of the agent, which may have a negative impact on performance. 
A more deeply-rooted weakness lies in the possible incomplete computation of the abstract state space. Indeed, since the latter is based on the episodes of the agent's training data, the abstraction of mutated episodes can be impossible (i.e, unseen abstract states are needed). 

\citeauthor{9794023}~\cite{9794023} also use a genetic algorithm, but aim at reducing the testing computation cost by assisting the search with surrogate models to avoid the expansive calls to the simulator.
More precisely, they reduce the number of fitness function computations (and so the simulation calls) by predicating the best test cases through cooperation between global and local searches. 
The results of the predictions of each iteration are then used to improve the surrogate models' accuracy.
Furthermore, to overcome the trade-off challenge of finding a balanced number of local surrogate models (accuracy versus performance), they introduce a clustering-based approach that generates one local surrogate model per cluster composed of test cases that belong to the same promising area.

\paragraph{RL-based Optimization.} \citeall{9712397} and \citeall{https://doi.org/10.48550/arxiv.2210.15432} turn the search problem into a RL task. The general idea is to train an agent to change in real-time the environment of the simulations towards situations where the policy under test exhibits faults. 
\citeauthor{9712397}~\cite{9712397} consider complex, highly configurable  environments for testing autonomous vehicles and learn a Deep Q-learning agent to find scenarios that maximise their collision. 
They compare the safety and current distances between the vehicle under test and its surrounding obstacles to estimate a collision probability (in a worst-case scenario) that is then used to define the reward function of the underlying MDP of the RL task. 
By linking the agent's rewards with the collision probability, their framework DeepCollision effectively trains the agent to guide the simulation towards collision-prone scenarios. 
On the other hand, \citeauthor{https://doi.org/10.48550/arxiv.2210.15432}~\cite{https://doi.org/10.48550/arxiv.2210.15432} check several safety requirements (i.e, the problem statement is a many-objective search). 
Consequently, their framework MORLOT considers a Q-table per safety requirement and the choice of every next action is based on the Q-table whose related safety property has not been violated yet and whose reward for the previously chosen action was the maximum. Regarding the reward function definitions, they are not based on the maximisation of the collision probability of the vehicle under test (as proposed by \citeauthor{9712397}~\cite{9712397}) but, rather, on the degree violation of the requirements. 
Those functions have thus to be defined for each safety property, as well as their respective maximum acceptable violation threshold.

Interestingly, by leveraging the same overall approach, these two works highlight its limitations and points of concern.
Indeed, we can note that they both define the MDP model of the RL task with context-dependent knowledge (e.g, changing weather or traffic conditions). 
More importantly, they also have to constrain the agent with hand-crafted, behavioral rules to ensure the consistency or realism of the simulations. 
Eventually, the two methodologies involve a significant number of parameters, whose definitions and performance impacts might be difficult to define and measure, respectively (more details in Section \ref{limitations}). 

\paragraph{Coverage-guided Search.}\label{coverageguidedsearch} \citeall{10.1145/3180155.3180220} detect erroneous behaviors of image input-based neural networks for driving autonomous cars as a neuron-coverage-guided greedy search.
At each step, the input state space is further explored with new synthetic images which are generated with MT.
The metamorphic operations to create those synthetic images involve weather condition changes (like adding fog or rain), and are assumed to keep the semantic of the original ones. As such, the metamorphic oracle checks if the outputs of the NN for the original and the new images are identical (given an error threshold). 
The search keeps track of the images which significantly increase the current neuron coverage to expand the input space. 
Even though the subsequent test cases do not eventually depict action scenarios like most of the works reviewed do, we decided to mention this work because we find the use of MT inside a guided search worth being mentioned. 
Regarding the scope and limitations of this approach (called DeepTest), one can note the use of neuron coverage. Indeed, in addition to recent concerns regarding the effectiveness of such a metric to guide search-based testing of NNs, it also restricts the approach to white-box testing setting. 
Besides, the metamorphic relations used in this work can misclassify correct behavior (since it is not guaranteed that the input transformations preserve the semantic of the images), meaning that the test suite generated is likely to include false positives.

\citeall{10.1145/3361566} also consider behavior inconsistencies testing of image input-based NNs.
This framework, DeepXplore, relies on a similar neuron-coverage-guided search (as DeepTest \cite{10.1145/3180155.3180220}). 
However, the oracle problem is not alleviated with MRs but, rather, with differential testing \cite{McKeeman1998DifferentialTF}, thus detecting erroneous behaviors when the NNs' outputs are not at all the same. 
Consequently, the search is formulated as a joint optimization problem to maximise both the neuron coverage and these output differences. 
Of course, DeepXplore is therefore only geared towards testing of multiple (white-box) neural networks.

\citeall{10.1145/3338906.3338954} study quantitative analysis of stateful RNN-based systems. 
They define an abstract model computation algorithm of the NN under test along with several quantitative indicators to enable adversarial attack detection and coverage-guided testing.
More precisely, they construct a Discrete-Time Markov chain by first applying state and transition abstractions on a set of concrete traces (of the RNN tested), and then computing transition probability distributions for each abstract state. 
The coverage metrics over the abstract model quantify its relative part exercised by a given concrete trace of the RNN, either state or transition wise. 
Those metrics are used to guide a test case search procedure similar to DeepTest \cite{10.1145/3180155.3180220}: new inputs are derived from the current test input with domain specific metamorphic transformations which are assumed to be semantically preservative. 
Consequently, faults are detected if the outputs of the RNN for the new inputs are not close enough to the initial one.
The benefit of guiding the search with coverage metrics over an abstract model of the RNN under test lets the framework consider the latter as a black-box. 
However, the subsequent weakness is that the guidance efficiency depends on the accuracy of the abstraction.
Indeed, in their experimental evaluation, the authors report that the level of abstraction granularity greatly impacts the resulting sensitivities of the coverage criteria. 

Eventually, \citeall{https://doi.org/10.48550/arxiv.2205.04887} propose an unusual, yet demanding search-based framework for safety testing of RL agents. The authors aim at generating what they call boundary test cases, that correspond to safety-critical situations. 
Precisely, a safety-critical situation is defined as a state of the environment in which an action can lead to the violation of a given safety property. 
The crucial difference with all the other search-based methodologies reviewed is that the framework does not look for those boundary states but, rather, retrieved the latter from the state space explored by an initial backtracking-based, depth-first search for a solution of the decision making problem.
The authors then define several test suites from those states. 
For instance, the ``simple test suite'' consists of all the states belonging to the paths that end in a boundary state. 
In conclusion, the key limitation of this methodology lies in the fact that it requires to solve the problem, and supposes that some boundary test cases will be found through the resolution computation. We further analyse that such a framework is difficult to apply to stochastic environments. 

\subsubsection{Fuzzing}
This subsection covers works whose testing procedure relies on fuzzing.
In general, fuzzing frameworks employ a pool of test case candidates (or seeds).
At each step, a seed is taken from the pool and used to create new test cases (with random transformations or mutation operations). 
The ones that successfully make the policy violate the testing objective(s) are added to the test suite. 
Those whose selection criterion value is high enough are inserted in the pool (or their associated seed). 
Additionally, some methodologies define specific strategies -- often called seed selection strategies -- to pick the most promising candidate from the pool, rather than using common random sampling for example.

\citeall{10.1145/3533767.3534388} consider initial simulation environments as seeds to test NN-based policies.
Test cases are generated by randomly mutating those initial environments with user-defined operations. 
For each test case executed, the framework computes the state sequence density of the execution trace of the agent.
The test cases with the highest density values are then used to feed the pool. 
The authors also propose to guide the choice of the next test case (to pick from the pool) with a metric called sensibility, which prioritises test cases that minimise the accumulated reward obtained by the agent (similarly to one of the fitness functions used in \cite{zolfagharian2023searchbased}).
Their intuition is that execution traces with low returns would guide the search towards situations where the agent is less robust and therefore, less safe. 
Their approach, which considers both the simulator and the policy as black-boxes, is implemented in a generic testing tool called MDPFuzz. 
However, the mutation operations as well as the test oracles (to detect errors among the execution traces) are input parameters and context-dependent.

\citeall{10.1145/3293882.3330579} study to what extent fuzzing frameworks are actually relevant for testing NNs in the first place. 
Like DeepTest, they consider image input-based NNs for decision making which are therefore not tested with simulation scenarios but, rather, with non-related test cases. 
They implement their approach in a tool called DeepHunter and extensively benchmark combinations of several seed selection and semantically preservative metamorphic mutation strategies, as well as well-known testing criteria to maintain the pool of candidates. 

Eventually, \citeall{Steinmetz_Fišer_Eniser_Ferber_Gros_Heim_Höller_Schuler_Wüstholz_Christakis_Hoffmann_2022} and \citeall{10.1145/3533767.3534392} investigate the bug confirmation problem for NN-based action policies.
More precisely, \citeauthor{Steinmetz_Fišer_Eniser_Ferber_Gros_Heim_Höller_Schuler_Wüstholz_Christakis_Hoffmann_2022}~\cite{Steinmetz_Fišer_Eniser_Ferber_Gros_Heim_Höller_Schuler_Wüstholz_Christakis_Hoffmann_2022} consider that a policy $\pi$ contains a bug if another policy $\pi'$ does better. The authors explore the use of heuristic functions used in classical AI Planning to automatically and efficiently compute test oracles (instead of computing such $\pi'$ policies). 
They also introduce a policy quality bias in the action selection of the random walks involved in the fuzzing framework. 

As for \citeauthor{10.1145/3533767.3534392}~\cite{10.1145/3533767.3534392}, they leverage a similar fuzzing, pool-based testing framework but rely on MT to automatically derive both the new environment to test the policy with and the associated test oracle. 
To do so, the authors design the metamorphic operations around state relaxation, a well-studied concept also taken from the AI Planning community. 
Their idea is that a relaxed version of a given environment should represent an easier problem than the original one.
Therefore, the agent's policy under test contains a bug if it solves the original problem but fails to solve its ``relaxed'' counterpart.

Some limitations regarding the works above are worth being mentioned. In MDPFuzz \cite{10.1145/3533767.3534388}, the consistency of mutation operations are checked arbitrarily, which is a similar weakness we found in DeepCollision \cite{9712397} and MORLOT \cite{https://doi.org/10.48550/arxiv.2210.15432}. 
Similarly, \citeauthor{10.1145/3293882.3330579}~\cite{10.1145/3293882.3330579} constrain the metamorphic mutation operations of DeepHunter with conservative parameters in order to best ensure the semantic equivalence with the original images. 
Still, they eventually assume that they are sufficient to keep the semantics of the mutated images. 
The approach proposed by \citeauthor{10.1145/3533767.3534392}~\cite{10.1145/3533767.3534392} is currently limited to invariant checking (i.e, non-temporal failure condition that must hold in every simulation state) and the state relaxation functions are input parameters (i.e, context-dependent). 
As for the work of \citeauthor{Steinmetz_Fišer_Eniser_Ferber_Gros_Heim_Höller_Schuler_Wüstholz_Christakis_Hoffmann_2022}~\cite{Steinmetz_Fišer_Eniser_Ferber_Gros_Heim_Höller_Schuler_Wüstholz_Christakis_Hoffmann_2022}, it currently comprises strong restrictions.
In particular, they consider decision making problems of classical AI Planning, where the dynamics of the environment are specified (i.e, white-box environment) and deterministic. 

\section{Limitation Summary}\label{limitations}

We identify redundant limitations as well as similar challenges among the contributions reviewed. 
We think that synthesising these findings can serve as guidelines for future researchers. 
Note that these limitations come on top of already recognized difficulties related to either V\&V practices or NNs, such as the high computational cost of some verification methods or the black-box nature of NN-based systems.

\paragraph{Context-dependent transformations.} Methodologies that involve mutation operations or environment transformations usually define the latter with respect to specific knowledge. As a result, the concerned frameworks, while being generically applicable, actually demand domain expertise. 
For instance, \cite{zolfagharian2023searchbased, 10.1145/3533767.3534388, https://doi.org/10.48550/arxiv.2210.15432} use mutation operations whose definitions are bound to the model of each decision making problem.

\paragraph{Assessing the consistency of the test cases is challenging.} Most of the works that mutate the test cases and/or change the environment of the simulations end up arbitrarily checking the consistency of the results. It is for example the case of MORLOT \cite{https://doi.org/10.48550/arxiv.2210.15432} and DeepCollision \cite{9712397}, which enforce the consistency of the online simulation modifications with hand-coded rules (e.g, the RL agents cannot modify the weather conditions ``too quickly''). Interestingly, MDPFuzz \cite{10.1145/3533767.3534388} mitigates this issue by generating the test cases before testing (instead of applying online modifications).

\paragraph{Hyperparameters.} One can distinguish two issues: methodologies with a significant number of parameters (often RL-based) and/or the ones for which the values of the parameters greatly impact the performance. For instance, the work of \citeauthor{zolfagharian2023searchbased}~\cite{zolfagharian2023searchbased} has an important number of parameters which, as a consequence, involves a great amount of effort to fine-tune. On the other hand, \citeauthor{9712397}~\cite{9712397} reported that too short time intervals between each action of their RL agent sometimes led simulations to chaotic driving situations.

\paragraph{Spurious results.} Some works rely on test oracles which might misclassify actually correct behavior. For example, \cite{10.1145/3293882.3330579, 10.1145/3238147.3238187, 10.1145/3180155.3180220} involve metamorphic operations that suppose the new inputs share the same semantic of their original counterparts, which is unfortunately not guaranteed (and difficult to assess). Similarly, the test oracles of MORLOT \cite{https://doi.org/10.48550/arxiv.2210.15432} depend on maximum acceptable degrees of violation, whose thresholds are left to hyperparameters.

\paragraph{Scope of applicability.} Some frameworks have significant applicability restrictions, that we see as noteworthy limitations. They can be on-purpose limitations (e.g, RNSVerify \cite{Akintunde_Kevorchian_Lomuscio_Pirovano_2019} specifically addresses the verification of RNN policies), strong testing assumptions (e.g, DeepXplore \cite{10.1145/3361566} test multiple, white-box NNs) or are due to the early nature of the methodologies (e.g, \citeauthor{10.1145/3533767.3534392}~\cite{10.1145/3533767.3534392} assume single non-temporal failure conditions and leave temporal properties as future work).   

\section{Future Research Direction}\label{conclusion}

Based on our literature review, we conclude this paper by elaborating on future research directions to guide and stimulate efforts for the V\&V of NN-based policies.

\paragraph{Extending existing works.} An important part of the papers analysed are pioneering works (e.g, \cite{10.1145/3533767.3534392, Steinmetz_Fišer_Eniser_Ferber_Gros_Heim_Höller_Schuler_Wüstholz_Christakis_Hoffmann_2022, Vinzent_Steinmetz_Hoffmann_2022}). 
As such, they have opened a new research area and raised questions which are now left to be addressed.
For instance, as mentioned previously, \citeauthor{10.1145/3533767.3534392}~\cite{10.1145/3533767.3534392} plan to extend their framework to temporal safety properties. 
The generalisation of the V\&V techniques could also be increased by relaxing their current testing setting requirements (e.g, white-box to black-box, deterministic to stochastic environments).

\paragraph{Refining existing works.} Research efforts are needed towards improving the methodologies themselves: whether it is performance enhancement (e.g, \citeauthor{9794023}~\cite{9794023} investigated surrogate models for many-objective search) or ease of applicability (e.g, enabling automated design of state relaxations in \cite{10.1145/3533767.3534392}). To that regard, \citeall{vinzent_neural_nodate} have recently enabled automated predicate abstraction computation (introduced in \cite{Vinzent_Steinmetz_Hoffmann_2022}) with counter-example guided abstraction refinement (CEGAR).

\paragraph{Combining different approaches/techniques.} Several research opportunities could consist in combining parts of existing works with each other to begin with. For instance, fuzzing frameworks can benefit from better guidance for their seed selection strategy and testing criterion. Furthermore, we find research interests in the investigation of different well-known software testing techniques than the ones the works reviewed opted for. For example, \citeauthor{Akintunde_Kevorchian_Lomuscio_Pirovano_2019}~\cite{Akintunde_Kevorchian_Lomuscio_Pirovano_2019} noted that they could have solved the unrolled FFNN verification task with SMT (instead of MILP).

\paragraph{Benchmarking and comparison studies.} Empirical studies are needed to better assess the applicability of some approaches. Moreover, this emerging research area lacks comparison evaluations. To that regard, we especially think that a scaling comparison between formal verification and testing methods could reveal possible limitations regarding the former.

\paragraph{Explainability of the policies tested.} Typical verification frameworks output SAT/UNSAT (possibly with a counterexample), which is not enough in the context of V\&V of intangible policies such as NN-based models. 
More informative results would help software engineers to understand and fix the non-compliant behaviors reported. Some of the works reviewed have already contributed to this end, like ProVe \cite{pmlr-v161-corsi21a} or DSMC \cite{Gros2022}, where a violation rate that quantifies the number of specification violations is introduced and a complete quality analysis of the neural network is provided, respectively. 
Similarly, finding fault-revealing test cases is far from being the only desirable feedback software engineers need. We argue that it is the very first step instead. 
Sharing this observation, we report early works among the ones reviewed. 
For example, MDPFuzz \cite{10.1145/3533767.3534388} investigates the visualisation of the distributions of the activated neurons by the fault-revealing states.
Additionally, we recall that \cite{Steinmetz_Fišer_Eniser_Ferber_Gros_Heim_Höller_Schuler_Wüstholz_Christakis_Hoffmann_2022, 10.1145/3533767.3534392} investigate bug confirmation, that captures avoidable failures among the defects of the policy under test.

\section*{Acknowledgements}
This work is funded by the Norwegian Ministry of Education and Research and part of the RESIST\_EA Inria-Simula associate team.

\bibliography{ref}

\begin{thebibliography}{42}
\providecommand{\natexlab}[1]{#1}
\providecommand{\url}[1]{\texttt{#1}}
\expandafter\ifx\csname urlstyle\endcsname\relax
  \providecommand{\doi}[1]{doi: #1}\else
  \providecommand{\doi}{doi: \begingroup \urlstyle{rm}\Url}\fi

\bibitem[fra(2014)]{frankish_ramsey_2014}
\emph{The Cambridge Handbook of Artificial Intelligence}.
\newblock Cambridge University Press, 2014.
\newblock \doi{10.1017/CBO9781139046855}.

\bibitem[Abiodun et~al.(2018)Abiodun, Jantan, Omolara, Dada, Mohamed, and
  Arshad]{Abiodun2018StateoftheartIA}
Oludare~Isaac Abiodun, Aman~Bin Jantan, Abiodun~Esther Omolara, Kemi~Victoria
  Dada, Nachaat Mohamed, and Humaira Arshad.
\newblock State-of-the-art in artificial neural network applications: A survey.
\newblock \emph{Heliyon}, 2018.

\bibitem[Akintunde et~al.(2019)Akintunde, Kevorchian, Lomuscio, and
  Pirovano]{Akintunde_Kevorchian_Lomuscio_Pirovano_2019}
Michael~E. Akintunde, Andreea Kevorchian, Alessio Lomuscio, and Edoardo
  Pirovano.
\newblock Verification of rnn-based neural agent-environment systems.
\newblock \emph{Proceedings of the AAAI Conference on Artificial Intelligence},
  2019.

\bibitem[Budde et~al.(2017)Budde, Dehnert, Hahn, Hartmanns, Junges, and
  Turrini]{10.1007/978-3-662-54580-5_9}
Carlos~E. Budde, Christian Dehnert, Ernst~Moritz Hahn, Arnd Hartmanns,
  Sebastian Junges, and Andrea Turrini.
\newblock Jani: Quantitative model and tool interaction.
\newblock In Axel Legay and Tiziana Margaria, editors, \emph{Tools and
  Algorithms for the Construction and Analysis of Systems}, 2017.

\bibitem[Chen et~al.(1998)Chen, Cheung, and Yiu]{Chen1998}
T.Y. Chen, S.C. Cheung, and S.M. Yiu.
\newblock Metamorphic {{Testing}}: {{A New Approach}} for {{Generating Next
  Test Cases}}.
\newblock Technical report, {Department of Computer Science, Hong Kong
  University of Science and Technology}, 1998.

\bibitem[Clarke and Schlingloff(2001)]{CLARKE20011635}
Edmund~M. Clarke and Bernd-Holger Schlingloff.
\newblock Chapter 24 - model checking.
\newblock In \emph{Handbook of Automated Reasoning}. North-Holland, 2001.

\bibitem[Corsi et~al.(2021)Corsi, Marchesini, and
  Farinelli]{pmlr-v161-corsi21a}
Davide Corsi, Enrico Marchesini, and Alessandro Farinelli.
\newblock Formal verification of neural networks for safety-critical tasks in
  deep reinforcement learning.
\newblock In \emph{Proceedings of the Thirty-Seventh Conference on Uncertainty
  in Artificial Intelligence}, 2021.

\bibitem[Corso et~al.(2022)Corso, Moss, Koren, Lee, and
  Kochenderfer]{10.1613/jair.1.12716}
Anthony Corso, Robert Moss, Mark Koren, Ritchie Lee, and Mykel Kochenderfer.
\newblock A survey of algorithms for black-box safety validation of
  cyber-physical systems.
\newblock \emph{J. Artif. Int. Res.}, 2022.

\bibitem[Du et~al.(2019)Du, Xie, Li, Ma, Liu, and
  Zhao]{10.1145/3338906.3338954}
Xiaoning Du, Xiaofei Xie, Yi~Li, Lei Ma, Yang Liu, and Jianjun Zhao.
\newblock Deepstellar: Model-based quantitative analysis of stateful deep
  learning systems.
\newblock In \emph{Proceedings of the 2019 27th ACM Joint Meeting on European
  Software Engineering Conference and Symposium on the Foundations of Software
  Engineering}, 2019.

\bibitem[Eniser et~al.(2022)Eniser, Gros, W\"{u}stholz, Hoffmann, and
  Christakis]{10.1145/3533767.3534392}
Hasan~Ferit Eniser, Timo~P. Gros, Valentin W\"{u}stholz, J\"{o}rg Hoffmann, and
  Maria Christakis.
\newblock Metamorphic relations via relaxations: An approach to obtain oracles
  for action-policy testing.
\newblock In \emph{Proceedings of the 31st ACM SIGSOFT International Symposium
  on Software Testing and Analysis}, 2022.

\bibitem[Ghosh et~al.(2018)Ghosh, Berkenkamp, Ranade, Qadeer, and
  Kapoor]{DBLP:journals/corr/abs-1802-08678}
Shromona Ghosh, Felix Berkenkamp, Gireeja Ranade, Shaz Qadeer, and Ashish
  Kapoor.
\newblock Verifying controllers against adversarial examples with bayesian
  optimization.
\newblock \emph{CoRR}, 2018.

\bibitem[Graf and Saidi(1997)]{10.1007/3-540-63166-6_10}
Susanne Graf and Hassen Saidi.
\newblock Construction of abstract state graphs with pvs.
\newblock In \emph{Computer Aided Verification}, 1997.

\bibitem[Gros et~al.(2021)Gros, H{\"o}ller, Hoffmann, Klauck, Meerkamp, and
  Wolf]{10.1007/978-3-030-85172-9_11}
Timo~P. Gros, Daniel H{\"o}ller, J{\"o}rg Hoffmann, Michaela Klauck, Hendrik
  Meerkamp, and Verena Wolf.
\newblock Dsmc evaluation stages: Fostering robust and safe behavior in deep
  reinforcement learning.
\newblock In Alessandro Abate and Andrea Marin, editors, \emph{Quantitative
  Evaluation of Systems}, 2021.

\bibitem[Gros et~al.(2022{\natexlab{a}})Gros, Hermanns, Hoffmann, Klauck,
  K{\"o}hl, and Wolf]{10.1007/978-3-031-13188-2_21}
Timo~P. Gros, Holger Hermanns, J{\"o}rg Hoffmann, Michaela Klauck,
  Maximilian~A. K{\"o}hl, and Verena Wolf.
\newblock Mogym: Using formal models for training and verifying decision-making
  agents.
\newblock In Sharon Shoham and Yakir Vizel, editors, \emph{Computer Aided
  Verification}, 2022{\natexlab{a}}.

\bibitem[Gros et~al.(2022{\natexlab{b}})Gros, Hermanns, Hoffmann, Klauck, and
  Steinmetz]{Gros2022}
Timo~P. Gros, Holger Hermanns, J{\"o}rg Hoffmann, Michaela Klauck, and Marcel
  Steinmetz.
\newblock Analyzing neural network behavior through deep statistical model
  checking.
\newblock \emph{International Journal on Software Tools for Technology
  Transfer}, 2022{\natexlab{b}}.

\bibitem[Haq et~al.(2022{\natexlab{a}})Haq, Shin, and Briand]{9794023}
Fitash~Ul Haq, Donghwan Shin, and Lionel Briand.
\newblock Efficient online testing for dnn-enabled systems using
  surrogate-assisted and many-objective optimization.
\newblock In \emph{2022 IEEE/ACM 44th International Conference on Software
  Engineering (ICSE)}, 2022{\natexlab{a}}.

\bibitem[Haq et~al.(2022{\natexlab{b}})Haq, Shin, and
  Briand]{https://doi.org/10.48550/arxiv.2210.15432}
Fitash~Ul Haq, Donghwan Shin, and Lionel Briand.
\newblock Many-objective reinforcement learning for online testing of
  dnn-enabled systems, 2022{\natexlab{b}}.

\bibitem[H{\'e}rault et~al.(2004)H{\'e}rault, Lassaigne, Magniette, and
  Peyronnet]{10.1007/978-3-540-24622-0_8}
Thomas H{\'e}rault, Richard Lassaigne, Fr{\'e}d{\'e}ric Magniette, and Sylvain
  Peyronnet.
\newblock Approximate probabilistic model checking.
\newblock In Bernhard Steffen and Giorgio Levi, editors, \emph{Verification,
  Model Checking, and Abstract Interpretation}, 2004.

\bibitem[Holland(1992)]{10.2307/24939139}
John~H. Holland.
\newblock Genetic algorithms.
\newblock \emph{Scientific American}, 1992.

\bibitem[Issakkimuthu et~al.(2018)Issakkimuthu, Fern, and
  Tadepalli]{Issakkimuthu_Fern_Tadepalli_2018}
Murugeswari Issakkimuthu, Alan Fern, and Prasad Tadepalli.
\newblock Training deep reactive policies for probabilistic planning problems.
\newblock \emph{Proceedings of the International Conference on Automated
  Planning and Scheduling}, 2018.

\bibitem[Karia and Srivastava(2020)]{DBLP:journals/corr/abs-2007-06702}
Rushang Karia and Siddharth Srivastava.
\newblock Learning generalized relational heuristic networks for model-agnostic
  planning.
\newblock \emph{CoRR}, 2020.

\bibitem[Lu et~al.(2023)Lu, Shi, Zhang, Zhang, Wang, Yue, and Ali]{9712397}
Chengjie Lu, Yize Shi, Huihui Zhang, Man Zhang, Tiexin Wang, Tao Yue, and
  Shaukat Ali.
\newblock Learning configurations of operating environment of autonomous
  vehicles to maximize their collisions.
\newblock \emph{IEEE Transactions on Software Engineering}, 2023.

\bibitem[Mandic and Chambers(2001)]{Mandic2001RecurrentNN}
Danilo~P. Mandic and Jonathon~A. Chambers.
\newblock Recurrent neural networks for prediction: Learning algorithms,
  architectures and stability.
\newblock 2001.

\bibitem[McKeeman(1998)]{McKeeman1998DifferentialTF}
William~M. McKeeman.
\newblock Differential testing for software.
\newblock \emph{Digit. Tech. J.}, 1998.

\bibitem[McMinn(2011)]{5954405}
Phil McMinn.
\newblock Search-based software testing: Past, present and future.
\newblock In \emph{2011 IEEE Fourth International Conference on Software
  Testing, Verification and Validation Workshops}, 2011.

\bibitem[Moore(1963)]{10.5555/904631}
Ramon~Edgar Moore.
\newblock \emph{Interval Arithmetic and Automatic Error Analysis in Digital
  Computing}.
\newblock PhD thesis, 1963.

\bibitem[Pang et~al.(2022)Pang, Yuan, and Wang]{10.1145/3533767.3534388}
Qi~Pang, Yuanyuan Yuan, and Shuai Wang.
\newblock Mdpfuzz: Testing models solving markov decision processes.
\newblock In \emph{Proceedings of the 31st ACM SIGSOFT International Symposium
  on Software Testing and Analysis}, 2022.

\bibitem[Pei et~al.(2019)Pei, Cao, Yang, and Jana]{10.1145/3361566}
Kexin Pei, Yinzhi Cao, Junfeng Yang, and Suman Jana.
\newblock Deepxplore: Automated whitebox testing of deep learning systems.
\newblock \emph{Commun. ACM}, 2019.

\bibitem[Silver et~al.(2016)Silver, Huang, Maddison, Guez, Sifre, Driessche,
  Schrittwieser, Antonoglou, Panneershelvam, Lanctot, Dieleman, Grewe, Nham,
  Kalchbrenner, Sutskever, Lillicrap, Leach, Kavukcuoglu, Graepel, and
  Hassabis]{article}
David Silver, Aja Huang, Christopher Maddison, Arthur Guez, Laurent Sifre,
  George Driessche, Julian Schrittwieser, Ioannis Antonoglou, Veda
  Panneershelvam, Marc Lanctot, Sander Dieleman, Dominik Grewe, John Nham, Nal
  Kalchbrenner, Ilya Sutskever, Timothy Lillicrap, Madeleine Leach, Koray
  Kavukcuoglu, Thore Graepel, and Demis Hassabis.
\newblock Mastering the game of go with deep neural networks and tree search.
\newblock \emph{Nature}, 2016.

\bibitem[Silver et~al.(2018)Silver, Hubert, Schrittwieser, Antonoglou, Lai,
  Guez, Lanctot, Sifre, Kumaran, Graepel, Lillicrap, Simonyan, and
  Hassabis]{doi:10.1126/science.aar6404}
David Silver, Thomas Hubert, Julian Schrittwieser, Ioannis Antonoglou, Matthew
  Lai, Arthur Guez, Marc Lanctot, Laurent Sifre, Dharshan Kumaran, Thore
  Graepel, Timothy Lillicrap, Karen Simonyan, and Demis Hassabis.
\newblock A general reinforcement learning algorithm that masters chess, shogi,
  and go through self-play.
\newblock \emph{Science}, 2018.

\bibitem[Steinmetz et~al.(2022)Steinmetz, Fišer, Eniser, Ferber, Gros, Heim,
  Höller, Schuler, Wüstholz, Christakis, and
  Hoffmann]{Steinmetz_Fišer_Eniser_Ferber_Gros_Heim_Höller_Schuler_Wüstholz_Christakis_Hoffmann_2022}
Marcel Steinmetz, Daniel Fišer, Hasan~Ferit Eniser, Patrick Ferber, Timo~P.
  Gros, Philippe Heim, Daniel Höller, Xandra Schuler, Valentin Wüstholz,
  Maria Christakis, and Jörg Hoffmann.
\newblock Debugging a policy: Automatic action-policy testing in ai planning.
\newblock \emph{Proceedings of the International Conference on Automated
  Planning and Scheduling}, 2022.

\bibitem[Tambon et~al.(2022)Tambon, Laberge, An, Nikanjam, Mindom, Pequignot,
  Khomh, Antoniol, Merlo, and Laviolette]{10.1007/s10515-022-00337-x}
Florian Tambon, Gabriel Laberge, Le~An, Amin Nikanjam, Paulina Stevia~Nouwou
  Mindom, Yann Pequignot, Foutse Khomh, Giulio Antoniol, Ettore Merlo, and
  Fran\c{c}ois Laviolette.
\newblock How to certify machine learning based safety-critical systems? a
  systematic literature review.
\newblock \emph{Automated Software Engg.}, 2022.

\bibitem[Tappler et~al.(2022)Tappler, Córdoba, Aichernig, and
  Könighofer]{https://doi.org/10.48550/arxiv.2205.04887}
Martin Tappler, Filip~Cano Córdoba, Bernhard~K. Aichernig, and Bettina
  Könighofer.
\newblock Search-based testing of reinforcement learning, 2022.

\bibitem[Tian et~al.(2018)Tian, Pei, Jana, and Ray]{10.1145/3180155.3180220}
Yuchi Tian, Kexin Pei, Suman Jana, and Baishakhi Ray.
\newblock Deeptest: Automated testing of deep-neural-network-driven autonomous
  cars.
\newblock In \emph{Proceedings of the 40th International Conference on Software
  Engineering}, 2018.

\bibitem[Toyer et~al.(2019)Toyer, Trevizan, Thi{\'{e}}baux, and
  Xie]{DBLP:journals/corr/abs-1908-01362}
Sam Toyer, Felipe~W. Trevizan, Sylvie Thi{\'{e}}baux, and Lexing Xie.
\newblock Asnets: Deep learning for generalised planning.
\newblock \emph{CoRR}, 2019.

\bibitem[Vinzent and Hoffmann(2022)]{vinzent_neural_nodate}
Marcel Vinzent and Joerg Hoffmann.
\newblock Neural {Policy} {Verification} via {Predicate} {Abstraction}:
  {CEGAR}.
\newblock page~9, 2022.

\bibitem[Vinzent et~al.(2022)Vinzent, Steinmetz, and
  Hoffmann]{Vinzent_Steinmetz_Hoffmann_2022}
Marcel Vinzent, Marcel Steinmetz, and Jörg Hoffmann.
\newblock Neural network action policy verification via predicate abstraction.
\newblock \emph{Proceedings of the International Conference on Automated
  Planning and Scheduling}, 2022.

\bibitem[Xiang et~al.(2018)Xiang, Tran, Rosenfeld, and Johnson]{8431048}
Weiming Xiang, Hoang-Dung Tran, Joel~A. Rosenfeld, and Taylor~T. Johnson.
\newblock Reachable set estimation and safety verification for piecewise linear
  systems with neural network controllers.
\newblock In \emph{2018 Annual American Control Conference (ACC)}, 2018.

\bibitem[Xie et~al.(2019)Xie, Ma, Juefei-Xu, Xue, Chen, Liu, Zhao, Li, Yin, and
  See]{10.1145/3293882.3330579}
Xiaofei Xie, Lei Ma, Felix Juefei-Xu, Minhui Xue, Hongxu Chen, Yang Liu,
  Jianjun Zhao, Bo~Li, Jianxiong Yin, and Simon See.
\newblock Deephunter: A coverage-guided fuzz testing framework for deep neural
  networks.
\newblock In \emph{Proceedings of the 28th ACM SIGSOFT International Symposium
  on Software Testing and Analysis}, 2019.

\bibitem[Zhang and Li(2020)]{ZHANG2020106296}
Jin Zhang and Jingyue Li.
\newblock Testing and verification of neural-network-based safety-critical
  control software: A systematic literature review.
\newblock \emph{Information and Software Technology}, 2020.

\bibitem[Zhang et~al.(2018)Zhang, Zhang, Zhang, Liu, and
  Khurshid]{10.1145/3238147.3238187}
Mengshi Zhang, Yuqun Zhang, Lingming Zhang, Cong Liu, and Sarfraz Khurshid.
\newblock Deeproad: Gan-based metamorphic testing and input validation
  framework for autonomous driving systems.
\newblock In \emph{Proceedings of the 33rd ACM/IEEE International Conference on
  Automated Software Engineering}, 2018.

\bibitem[Zolfagharian et~al.(2023)Zolfagharian, Abdellatif, Briand,
  Bagherzadeh, and S]{zolfagharian2023searchbased}
Amirhossein Zolfagharian, Manel Abdellatif, Lionel Briand, Mojtaba Bagherzadeh,
  and Ramesh S.
\newblock A search-based testing approach for deep reinforcement learning
  agents, 2023.

\end{thebibliography}
\bibliographystyle{plainnat}

\end{document}